\documentclass[sigconf]{acmart}
\usepackage{graphicx}
\usepackage[ruled]{algorithm2e}
\usepackage[utf8]{inputenc}
\usepackage{subfigure}
\usepackage{multirow}
\usepackage{tabulary}
\usepackage{balance}
\AtBeginDocument{%
  \providecommand\BibTeX{{%
    \normalfont B\kern-0.5em{\scshape i\kern-0.25em b}\kern-0.8em\TeX}}}

\copyrightyear{2020}
\acmYear{2020}
\setcopyright{acmcopyright}\acmConference[SIGIR '20]{Proceedings of the 43rd
International ACM SIGIR Conference on Research and Development in Information
Retrieval}{July 25--30, 2020}{Virtual Event, China}
\acmBooktitle{Proceedings of the 43rd International ACM SIGIR Conference on
Research and Development in Information Retrieval (SIGIR '20), July 25--30, 2020,
Virtual Event, China}
\acmPrice{15.00}
\acmDOI{10.1145/3397271.3401165}
\acmISBN{978-1-4503-8016-4/20/07}

\begin{document}

\title{GCN-Based User Representation Learning for Unifying Robust Recommendation and Fraudster Detection}
\fancyhead{}
\author{Shijie Zhang}
\affiliation{%
  \institution{The University of Queensland}}
\email{shijie.zhang@uq.edu.au}

\author{Hongzhi Yin}
\authornote{Corresponding author; contributing equally with the first author.}
\affiliation{%
  \institution{The University of Queensland}}
\email{h.yin1@uq.edu.au}
\author{Tong Chen}
\affiliation{%
  \institution{The University of Queensland}}
\email{tong.chen@uq.edu.au}
\author{Quoc Viet Nguyen Hung}
\affiliation{%
  \institution{Griffith University}}
\email{quocviethung.nguyen@griffith.edu.au}
\author{Zi Huang}
\affiliation{%
  \institution{The University of Queensland}}
\email{huang@itee@uq.edu.au}
\author{Lizhen Cui}
\affiliation{%
  \institution{Shandong University}}
\email{clz@sdu.edu.cn}

\begin{abstract}
In recent years, recommender system has become an indispensable function in all e-commerce platforms. The review rating data for a recommender system typically comes from open platforms, which may attract a group of malicious users to deliberately insert fake feedback in an attempt to bias the recommender system to their favour. The presence of such attacks may violate modeling assumptions that high-quality data is always available and these data truly reflect users' interests and preferences. Therefore, it is of great practical significance to construct a robust recommender system that is able to generate stable recommendations even in the presence of shilling attacks. In this paper, we propose GraphRfi - a GCN-based user representation learning framework to perform robust recommendation and fraudster detection in a unified way. In its end-to-end learning process, the probability of a user being identified as a fraudster in the fraudster detection component automatically determines the contribution of this user's rating data in the recommendation component; while the prediction error outputted in the recommendation component acts as an important feature in the fraudster detection component. Thus, these two components can mutually enhance each other. Extensive experiments have been conducted and the experimental results show the superiority of our GraphRfi in the two tasks - robust rating prediction and fraudster detection. Furthermore, the proposed GraphRfi is validated to be more robust to the various types of shilling attacks over the state-of-the-art recommender systems.  
\end{abstract}

%
%


%
\maketitle
\vspace{-1.5em}
\section{INTRODUCTION}
With the explosive growth of e-commerce, more and more customers prefer shopping online via various e-commerce sites (e.g., Amazon and Yelp)~\cite{li2019capsule,zhang2019deep}. As users are exposed to a wider range of online and products, it becomes increasingly challenging for users to choose the right products within the limited time. A successful mechanism to help customers alleviate such information overload is the recommender system~\cite{koren2015advances, yin2019social, wu2016collaborative}. The main idea is to predict the ratings of a set of unrated items for a user based on her/his historical behaviours, and then the personalized recommendations can be selected from items with top predicted ratings. 

In order to provide good user experience, e-commerce platforms have a strong desire to keep their recommendation results highly accurate. In this context, recommendation accuracy means that the predicted ratings should be as close as possible to their true rating values, and inaccurate rating predictions will lead to unsatisfying recommendations and decreased competitiveness against rival companies. Currently, with the assumption that high-quality user rating data is available (i.e., the collected user rating data can truly represent the user preference), most of the recommender systems can achieve the aforementioned goal~\cite{aggarwal2016recommender,zhang2019deep,yin2018joint,xie2016learning}. Among various recommendation techniques, the most popular and efficient one is collaborative filtering (CF), which can be categorized into memory-based and model-based approaches. User-based~\cite{zhao2010user} and item-based CF~\cite{sarwar2001item} are memory-based algorithms assuming that similar users share similar interests or similar items have similar characteristics. PMF~\cite{mnih2008probabilistic} and MF~\cite{koren2009matrix} are representatives of model-based CF, which decompose the user-item interaction matrix into the product of two lower dimensional matrices. More recently, deep learning (DL) techniques have been revolutionizing the recommendation architectures dramatically and opening up more opportunities to improve the performance of recommender system~\cite{zhang2019deep, guo2017deepfm,zhang2019inferring,chen2020social}. For example, Berg et al.~\cite{berg2017graph} and Fan et al.~\cite{wen2019www} adopt graph convolutional networks that account for external user/item information for boosting the recommendation performance.

However, it is reported in many studies~\cite{mehta2007robust, aggarwal2016recommender,jannach2011recommender} that recommender systems have attracted many attackers to submit incorrect feedback about items to trap or manipulate the recommendation systems. Due to financial incentives, a group of imposters may try to game the system by posting fake feedback and unfair ratings to either promote or defame their targeted items~\cite{DBLP:conf/icwsm/MukherjeeV0G13}. Such users are called fraudsters~\cite{kumar2018rev2,dong2018opinion} and such activities to manipulate recommendation results are termed shilling attack~\cite{wu2012hysad,mehta2007unsupervised} which can cause ill-posed recommendations and harm the fairness of e-commerce markets. A real example is that, in 2015, Samsung was alleged by Taiwan's Fair Trade Commission to have hired students to post negative comments about HTC phones, and was confronted with a fine of 25 million Taiwanese dollars~\cite{chen2015opinion}. In the presence of shilling attacks, the widely-used CF-based recommendation methods are subject to different levels of performance drop. This is because the models are designed without the awareness of untruthful ratings, and the user/item representations are learned from such misinformation, thus being biased and unreliable. 
To resist the shilling attack on the recommender systems, there are two pathways in contemporary research work. One is to increase the robustness of recommendation models with the existence of shilling attacks, while the other is to directly detect and block fraudsters from the data. These two methods have been studied separately, and the potential of integrating them into a unified framework to make most of their complementary merits remains largely unexplored. The study of robust recommender systems mainly focuses on introducing robust statistical methods such as M-estimators~\cite{huber2004robust} to the optimization function of matrix factorization~\cite{mehta2007robust}. However, many invalidated assumptions are made in this method, and it is also hard to exploit and integrate the rich side information of users. As the second solution, fraudster detection classifiers~\cite{o2004evaluation,mehta2009unsupervised} are proposed to identify fraudsters from rating datasets, and then the recommendations can be performed after removing the detected fraudsters from the datasets. However, due to the pursuit of high classification recall rates, the removal of fraudsters is a mistake-prone process, in which genuine users might be labeled as fraudsters and thus removed, resulting in counter-productive effects~\cite{aggarwal2016recommender}. Accordingly, the proper use of fraudster detection is a pivotal part in the design of a robust recommender system. These limitations motivate us to design a novel model framework that meets the following two properties. Firstly, it should be able to fully utilize the available users' reliability features (e.g., the entropy of ratings and positive/negative rating distributions~\cite{dong2018opinion, mukherjee2013spotting}) to improve the robustness of the model when it is exposed to noisy data caused by shilling attacks. Secondly, it should be a unified, end-to-end trainable model that learns a universal representation for both robust recommendation and fraudster detection. 

In this paper, we propose an end-to-end GCN-based user representation learning (\textbf{GraphRfi}) framework that consists of two main components respectively focusing on robust recommendation and fraudster detection. These two components are coupled with each other and mutually enhance each other in the learning process. To fully exploit and integrate the rich side information of users, we extend the Graph Convolutional Network (GCN)~\cite{defferrard2016convolutional} to model the user-item interactions and predict user ratings in the recommendation component. GCN naturally integrates the node features as well as the topological structure, and thus it can be powerful in representation learning. We first predefine a variety of features that are indicative for detecting fraudsters, which are then used to initialize user nodes in the GCN. By iteratively aggregating predefined feature information from local user-item graph neighborhoods using GCN, the representation of users can capture both their preference and reliability information in a unified way. From the perspective of Cognitive Psychology, a normal user’s behaviours should be consistent and predictable~\cite{white2015consistency}. Therefore, if a user’s obvious rating is largely deviated from her/his predicted rating by an accurate recommender model, this user is most likely to be a fraudster. Therefore, we take the aggregated rating prediction error of each user as an important input feature for the fraudster detection. In the fraudster detection component, we adopt the neural random forests~\cite{kontschieder2015deep} due to its excellent performance in a wide range of classification tasks. The probability of a user being classified as a fraudster is taken as a weight to control the contribution of this user's rating data to the loss function of the recommendation component. If a user has a high probability of being a fraudster, GraphRfi can reduce the impact of this user's generated rating on the trained recommender model. As a result, these two components are closely hinged and will exert mutually.

Overall, the primary contributions of our research are summarized below.
\begin{itemize}
    \item To the best of our knowledge, we are the first to investigate the interdependent and mutually beneficial relationship between the robust recommendation task and the fraudster detection task. 
    \item We propose a novel end-to-end user representation learning framework with GCN and neural random forest as main building blocks, namely GraphRfi,  to capture and represent  both user preference and reliability  in a unified way.  
    \item Extensive experiments are conducted to evaluate the performance of GraphRfi, and the experimental results show its superiority in terms of recommendation accuracy, recommendation robustness, and fraudster detection accuracy. 
\end{itemize}
\section{GRAPHRFI: THE MODEL}
\begin{figure*}
\centering
\includegraphics[scale=0.60]{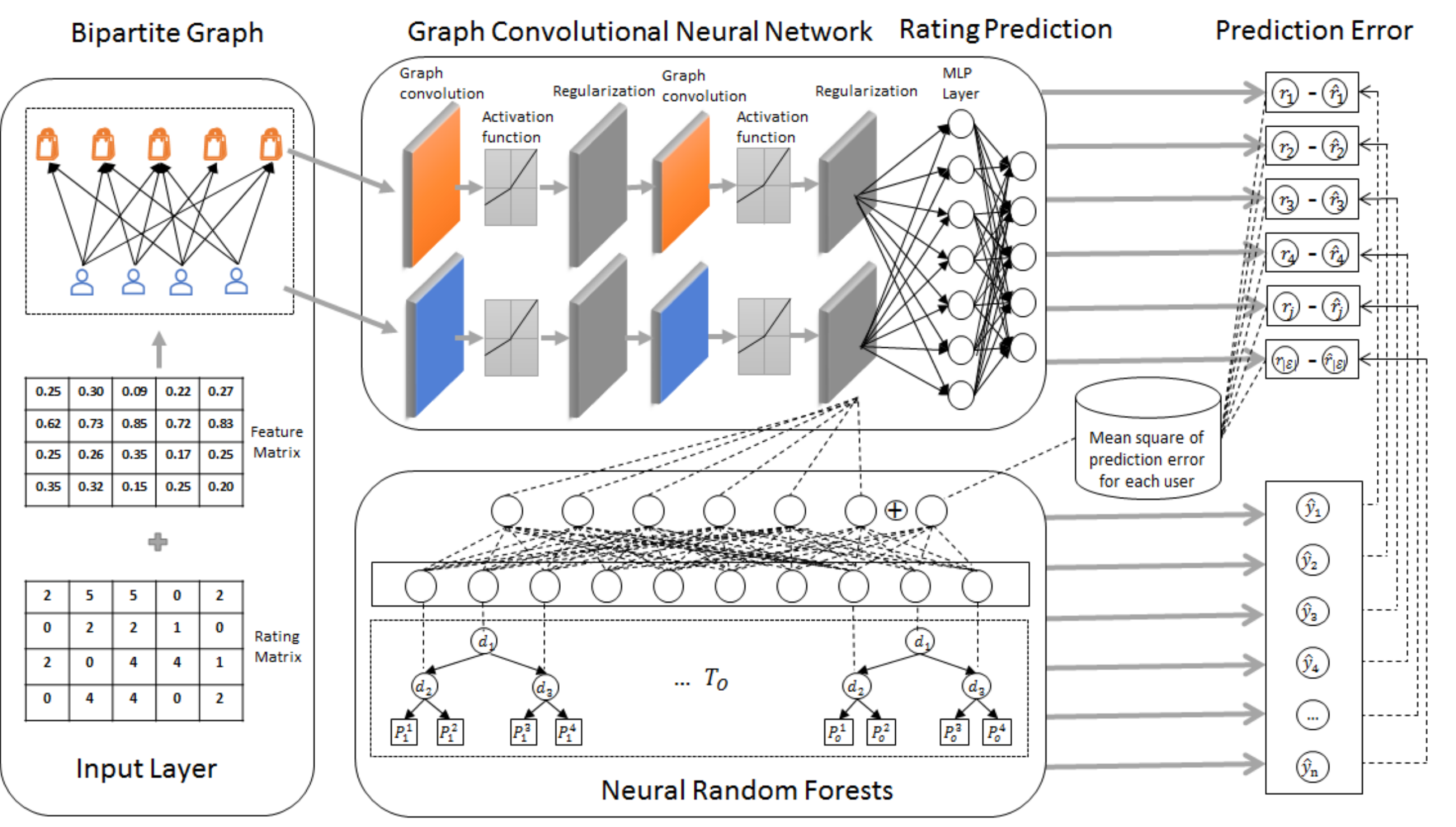}
\vspace{-0.8em}
\caption{The overview of GraphRfi}
\label{fig:frame}
\vspace{-1.4em}
\end{figure*}
In this section, we first mathematically formulate the two key tasks: robust rating prediction and fraudster detection, and then present the technical details of our proposed model GraphRfi. Note that in the description below, all vectors and matrices are respectively denoted with bold small letters and bold capital letters, and all sets are calligraphic uppercase letters.
\vspace{-0.5em}
\subsection{Preliminaries}
\label{st:features}
In a rating system, users' behavioural footprints are very indicative of their own characteristics. Thus, for user $u$, we first construct a dense feature vector $\mathbf{x}_u \in \mathbb{R}^{b}$ with each element representing a predefined statistical property of $u$'s behaviors. Each feature $x_{ui}\in \mathbf{x}_u$ is described in Table~\ref{tab:ff}, where the statistics are based on mainstream 5-star rating systems such as Amazon and Yelp. As shown in~\cite{dong2018opinion,mukherjee2013spotting}, the behavioral features can provide important signals for identifying fraudsters, and thus they are exploited and integrated to learn latent user representations. We use $\mathbf{X}$ to denote the feature matrix of all users. 

\begin{table}[h]
\vspace{-0.5em}
  \scalebox{0.9}{%
  \begin{tabular}{p{9cm}}
    \toprule
    \midrule
\textbf{- Number of rated products}\\
\textbf{- Length of username}\\
\textbf{- Number and ratio of each rating level given by a user}\\
\textbf{- Ratio of positive and negative ratings}: The proportions of high ratings (4 and 5) and low ratings (1 and 2) of a user.\\
\textbf{- Entropy of ratings}: As a measure of skewness in ratings, it can be calculated as $-\sum_{\forall r}P_{r}log P_{r}$, where $P_{r}$ is the proportion that a user gives the rating of $r$.\\
\textbf{- Total number of helpful and unhelpful votes a user gets}\\
\textbf{- The ratio and mean of helpful and unhelpful votes}\\
\textbf{- Median, min, and max number of helpful and unhelpful votes}\\
\textbf{- Day gap}: The number of days between a user's first and last rating.\\
\textbf{- Time entropy}: It measures the skewness in user's rating time (by year). It is calculated as $-\sum_{j=1}^{\tau}t_{j}log t_{j}$, where $\tau$ is the time gap between a user's first and last ratings, and $t_{j}$ is the proportion of ratings this user has generated in year $j$.\\
\textbf{- Same date indicator}: It indicates whether a user's first and last comments are on the same date. The value is 1 if yes, and 0 otherwise.\\
\textbf{- Median, min, max, and average of ratings}: These are further statistics to help identify fraudsters with the intuition that fraudsters are likely to give extreme ratings in order to demote/promote products.\\
\textbf{- Feedback summary length}: The number of words in the feedback.\\
\textbf{- Review text sentiment}: We use 1, 0, -1 to respectively represent positive, neutral and negative sentiment of a user's all reviews.\\
    \midrule
   \bottomrule
\end{tabular}}
\caption{Feature Description}
\label{tab:ff}
\vspace{-2.5em}
\end{table}
\textit{Task 1.} \textbf{Robust Rating Prediction:} We define a rating graph as a weighted bipartite graph $\mathcal{G} = (\mathcal{U} \cup \mathcal{V}, \mathcal{E})$. Here, $\mathcal{U} = \{u_{1}, u_{2}, ..., u_{n} \}$, $\mathcal{V} = \{v_{1}, v_{2}, ..., v_{m}\}$ represent the sets of $n$ users and $m$ items, respectively. A directed edge $(u, v, r_{uv}) \in \mathcal{E}$ means that user $u$ has rated item $v$ with score $r_{uv}$, and each user is associated with behavioral side information $\textbf{x}_u$. We use $\mathbf{R} \in \mathbb{R}^{n \times m}$ to denote the user-item rating matrix. Note that there are also malicious users (i.e., fraudsters) $u' \in \mathcal{U}$ as well as untruthful ratings (i.e., shilling attacks) $r_{u'v} \in \textbf{R}$ made by fraudsters. With the existence of fraudsters and shilling attacks, given the rating graph $\mathcal{G}$ with user feature vectors $\textbf{x}_u$, we aim to predict the true rating that a non-malicious user would give to an item that he/she has not rated in the past. 

\textit{Task 2.} \textbf{Fraudster Detection:} Given a set of users $\mathcal{U}$, with their rating data $\mathbf{R}$ and feature matrix $\mathbf{X}$, this task aims to classify these users into fraudsters and genuine users.  
\vspace{-1.5em}
\subsection{Overview of GraphRfi Framework}
The GraphRfi framework is shown in Figure 1, which consists of two key components to perform robust rating prediction and fraudster detection respectively.  In these two components, we adopt the graph convolutional network (GCN)~\cite{defferrard2016convolutional} and neural random forest (NRF)~\cite{kontschieder2015deep} as the building blocks, because the GCN~\cite{defferrard2016convolutional} is capable of fully exploiting local structure information of the rating graph $\mathcal{G}$ and the user side information $x_u$ to capture both user preferences and reliability in a unified way, and the NRF has achieved outstanding performance in a variety of classification tasks. Note that these two components are closely coupled and mutually enhanced with each other, making GraphRfi an end-to-end learning framework. First, these two components share the same user embeddings. Second, the probability of a user being classified as a fraudster by the NRF component is taken as a weight to control the contribution of this user's generated rating data in the GCN component, while the aggregated prediction error of the user's ratings is taken as an important feature for the fraudster detection component (i.e., the NRF). In what follows, we will introduce each component in details.
\subsection{GCN-based Rating Prediction Component}

\label{st:gcn}
Given a rating graph $\mathcal{G} = (\mathcal{U} \cup \mathcal{V}, \mathcal{E})$, we assume that each user $u \in \mathcal{U}$ is associated with  a feature vector $\mathbf{x}_{u}$. As the rating graph is a weighted bipartite graph with multiple information sources, the main challenge is how to collaboratively leverage both the node features as well as the graph's structural information to generate high-quality embeddings. Therefore, we adopt GCN as the main building block in this component to simultaneously capture both the topological neighborhood structure and the side information of each node in a natural way.  

We start describing the GCN architecture from the user side. Note that we use $\textbf{z}$ to represent vectors of both users and items in the same embedding space, where each user node and item node are indexed by subscripts $u$ and $v$ respectively.
Each user's behavioral feature vector $\textbf{x}_u$ is taken as the initial embedding, i.e., $\textbf{z}_u^{0} = \textbf{x}_u$,
while each item embedding $\textbf{z}_v^0$ is initialized using randomized values. The interactions between users and items are treated as a form of information passing, where vector-valued node messages are passed through user-item edges. In each forward iteration, we transform the information passed from user $u$'s interacted items into dense representations using deep neural networks, and merge all the received information with the original user embedding to update the user representation.

Specifically, we introduce the representation of message $\mathbf{h}_{v}$ passed from item $v \in \mathcal{N}(u)$, where $\mathcal{N}(u)$ is the set of items rated by $u$. As the observed ratings reflect users' explicit preferences towards items, different rating score (e.g., rating stars) represents different degrees of preferences. To distinguish the different rating scores, for each explicit rating score $r$, we introduce an embedding vector $\mathbf{e}_{r} \in \mathbb{R}^{e}$ as its latent representation. Then, the message $\mathbf{h}_{v}$ can be modelled by combining the item representation $\mathbf{z}_{v}$ with the rating representation $\mathbf{e}_{r_{uv}}$ as follows:
\begin{equation}\label{eq:information}
    \mathbf{h}_{v} = g(\mathbf{z}_{v} \oplus \mathbf{e}_{r_{uv}}) 
\end{equation}
where $\oplus$ denotes the concatenation of two vectors, and $g(\cdot)$ is a multi-layer perceptron (MLP) with $l$ hidden layers which helps encode the information of both the item and associated user sentiment into a compact vector representation. Then, upon receiving the edge-specific messages from $u$'s interacted items, an information aggregation operation is performed to fuse all received information into the current user representation $\textbf{z}_u$, which results in an updated user representation $\textbf{z}_u^{new}$. Before we proceed, we let $\mathcal{I}(u) = \mathcal{N}(u)\cup\{u\}$, where $\mathcal{I}(u)$ contains both the user node $u$ and her/his rated items $\mathcal{N}(u)$. We term $\mathcal{I}(u)$ the \textit{neighbor set} for $u$. Then, mathematically, the information aggregation is computed as below:
\begin{equation}\label{eq:convolution}
    \mathbf{z}_{u}^{new} = \sigma(\mathbf{W} \cdot Agg(\{\mathbf{h}_{k}, \forall k \in \mathcal{I}(u)\})+ \mathbf{b})
\end{equation}
where $\mathbf{h}_{k}$ is the information associated with each node within $\mathcal{I}(u)$. In the graph convolutional network, $\sigma$ denotes the rectified linear unit ($ReLU(\cdot)$) for non-linear activation, while $\mathbf{W}$ and $\mathbf{b}$ are corresponding trainable weight matrix and bias vector. Importantly, $Agg(\cdot)$ is the aggregation function which models the incoming information passed to user $u$ by combining the information of all $\mathbf{h}_{k}$ into a unified vector representation. Due to the fact that different interacted items carry different characteristics and tend to have varied contributions to learning the user's representation, we propose to incorporate the attention mechanism \cite{velivckovic2017graph} into the design of $Agg(\cdot)$. Specifically, for the input neighbor set $\{\textbf{h}_k, \forall k \in \mathcal{I}(u)\}$, we define $Agg(\{\textbf{h}_k, \forall k \in \mathcal{I}(u)\})$ as:
\begin{equation}
    Agg(\{\textbf{h}_{k}, \forall k \in \mathcal{I}(u)\}) = \sum_{k\in \mathcal{I}(u)}\alpha_{uk}\mathbf{h}_{k}
\end{equation}
where $\alpha_{uk}$ denotes the attention weight quantifying the importance of each neighbor node to user $u$ during information aggregation. In order to obtain sufficient expressiveness when propagating lower-level user features into higher-level features, we parameterize the attention weight $\alpha_{uk}$ with a two-layer neural network, which takes the message $\mathbf{h}_{k}$ and the target user's embedding $\mathbf{z}_{u}$ as its input:
\begin{equation}
    a_{uk} = {\mathbf{w}_{2}^{\top}} \cdot \sigma(\mathbf{W}_{1} \cdot [\mathbf{h}_{k} \oplus \mathbf{z}_{u}] + \mathbf{b}_{1}) + b_{2}
\end{equation}
The attention score $a_{uk}$ indicates the feature-level importance of neighbor node $k\in \mathcal{I}(u)$ to user $u$. Afterwards, we obtain each final attention weight $\alpha_{uk}$ by normalizing all the scores using softmax:
\begin{equation}
      \alpha_{uk} = \frac{exp(a_{uk})}{\sum_{k'\in \mathcal{I}(u)} exp(a_{uk'})}
\end{equation}

With the aforementioned operations, we can correspondingly model the latent representations of items. Essentially, we interchange the roles of item nodes and user nodes described in Eq.(\ref{eq:information}) and Eq.(\ref{eq:convolution}), and leverage the information passed from an item's connected users (denoted by $\mathcal{N}(v)$) to learn the item embedding $\textbf{z}_v$. Formally, the computation for an item node $v$ requires us to firstly compute the message passed from user $u \in \mathcal{U}$, represented by $\textbf{h}_{u}$:
\begin{equation}
    \mathbf{h}_{u} = g(\mathbf{z}_{u} \oplus \mathbf{e}_{r_{uv}}) 
\end{equation}
and use it for updating item representation $\textbf{z}_v^{new}$:
\begin{equation}
    \mathbf{z}_{v}^{new} = \sigma(\mathbf{W} \cdot Agg(\{\mathbf{h}_{q}, \forall q \in \mathcal{I}(v)\})+ \mathbf{b})
\end{equation}
where $\mathcal{I}(v) = \mathcal{N}(v)\cup\{v\}$ is the neighbor set for item $v$.

We use Algorithm 1 to illustrate a graph convolution step for all user and item nodes in $\mathcal{G}$. As such, we can acquire user and item representations, i.e., $\textbf{z}_u^{new}$ and $\textbf{z}_v^{new}$, that comprehensively incorporate information of both the user/item nodes themselves and all item/user nodes connected to them. Note that we only consider directly observed user-item interactions, i.g., first-order connections in our GCN framework. It is also worth mentioning that the calculation process for both users and items share the same network structure and trainable parameters.
 \begin{algorithm}
            \caption{A Full Graph Convolution Step}
            \KwIn{Graph $\mathcal{G} = (\mathcal{U} \cup \mathcal{V}, \mathcal{E})$, network parameters $\{ \mathbf{W}, \mathbf{b}, \mathbf{W}_{1}, \mathbf{w}_{2}, \mathbf{b}_{1}, b_{2}\}$, user current embedding $\mathbf{z}_{u}$, item current embedding $\mathbf{z}_{v}$}
            \KwOut{Updated user and item representations $\textbf{z}_{u}^{new}$ and $\textbf{z}_v^{new}$}
            
                \For{$u\in \mathcal{U}$}
                {
                \For{$v\in \mathcal{N}(u)$}
                {
                $\mathbf{h}_{v} \leftarrow g(\mathbf{z}_{v} \oplus \mathbf{e}_{r_{uv}})$\;
                }             
                $\mathbf{z}_{u}^{new} \leftarrow \sigma(\mathbf{W} \cdot Agg(\{\mathbf{h}_{k}, \forall k \in \mathcal{I}(u)\})+ \mathbf{b})$\;
                }                
                \For{$v\in \mathcal{V}$}
                {
                \For{$u\in \mathcal{N}(v)$}
                {
                $\mathbf{h}_{u} \leftarrow g(\mathbf{z}_{u} \oplus \mathbf{e}_{r_{uv}})$\;
                }             
                $\mathbf{z}_{v}^{new} \leftarrow \sigma(\mathbf{W} \cdot Agg(\{\mathbf{h}_{q}, \forall q \in \mathcal{I}(v)\})+ \mathbf{b})$\;
                }
        
    \end{algorithm}

With the user and item representations learned from the rating graph, we can subsequently predict the rating score $r_{uv}$ for a given user-item pair. To achieve this, we first concatenate their latent representations, and use another $l$-layer MLP $g'(\cdot)$ coupled with a projection layer for rating regression:
\begin{equation}
\label{sec:rp}
    \hat{r}_{uv} = \textbf{w}_{project}^{\top} g'(\textbf{z}_u^{new} \oplus \textbf{z}_v^{new}) 
\end{equation}
where $\hat{r}_{uv}$ denotes the predicted rating from user $u$ to item $v$. By measuring the error of predicted $\hat{r}_{uv}$ against the ground truth rating $r_{uv}$, we can optimize the MLP-based regressor for rating prediction.
\vspace{-1em}
\subsection{NRF-based Fraudster Detection Component}
\label{st:ffd}
In this component, we design a supervised classifier to detect fraudsters, where the set of user identity labels denotes as $\mathcal{Y}$ ($y_u \in \mathcal{Y} = 1$ if $u$ is a fraudster, and $y_u = 0$ if $u$ is a genuine user). A critical innovation in our approach is that, for each user $u$, we calculate the mean square of all rating prediction errors on the interacted items $\mathcal{N}(u)$, as an additional user feature for distinguishing fraudsters. Specifically, the additional feature $error_u$ is calculated below:
\begin{equation}
error_u = \frac{1}{|\mathcal{N}(u)|}\sum_{\forall v \in \mathcal{N}(u)}(|r_{uv} - \hat{r}_{uv}|^2)	
\end{equation}
Then, we append $error_u$ to the corresponding user embedding $\textbf{z}_{u}^{new}$ learned in Section~\ref{st:gcn} for classification. We denote the enhanced user feature vector as $\textbf{z}_u'=\textbf{z}_{u}^{new} \oplus error_u$. The reason for incorporating this feature is that normal users are more stable and predictable than the malicious users from the perspective of cognitive psychology \cite{white2015consistency}. Therefore, if a user's ratings largely deviate from the predicted ones, $error_u$ can provide a strong signal that this user is likely to be a fraudster. 

To make GraphRfi end-to-end trainable, each network component should be differentiable. As a result, motivated by \cite{kontschieder2015deep}, we design a neural network-based variant of the random forest, namely neural random forest (NRF). Compared with conventional random forest, NRF is a stochastic and differentiable extension to it, thus ensuring both the model expressiveness and end-to-end compatibility. Specifically, in NRF, each user will be assigned a probability, making the decision tree fully differentiable. Firstly, with the concatenated feature $\textbf{z}_u'$, we let it go through a fully connected layer to obtain a dense representation $\mathbf{z}^{*}_u$:
\begin{equation}
    \mathbf{z}_u^{*} = Sigmoid(\mathbf{W}_{Z'}\textbf{z}_u' + \mathbf{b}_{Z'})
\end{equation}
where $\mathbf{W}_{Z'}$, $\mathbf{b}_{Z'}$ are the weight and bias terms. Then, we will introduce how NRF performs fraudster classification with input $\mathbf{z}_u^*$.

Suppose we have $O$ decision trees and all the trees follow the standard binary tree structure. Each tree $T_{o}$ ($o \in [1,O]$) is a classifier consisting of two types of nodes, namely prediction nodes $p \in \mathcal{P}_o$ and decision nodes $d \in \mathcal{D}_o$. Prediction nodes are the terminal nodes (leaf nodes) of the tree and each prediction node $p$ holds a probability distribution $\pi_{p}$ over the label $y\in\{0,1\}$, i.e., $\pi_{p} = [\pi_{p_0}=\mathbb{P}(y=0),\pi_{p_1}=\mathbb{P}(y=1)]$. Each decision node (i.e., non-leaf node) $d \in \mathcal{D}$ represents a decision function $f_{d}(\mathbf{z}_u^* ; \Theta): \mathbf{z}_u^{*} \mapsto [0,1]$ parameterized by $\Theta$, which is responsible for determining whether the received input $\mathbf{z}_u^{*}$ will be sent to its left or right subtree \cite{kontschieder2015deep}. For each decision node, the decision function is defined as follows:
\begin{equation}
    f_{d}(\mathbf{z}_u^{*};\Theta) = Sigmoid(\mathbf{w}_{d}^{\top}\mathbf{z}_{u}^{*})
\end{equation}
which projects the input $\mathbf{z}_u^{*}$ into a scalar, and then produces a probability approximating the ground truth label.

Accordingly, for each tree $T_o$, the probability of having user $u$ to be classified as label $y$ would be:
\begin{equation}
    \mathbb{P}_{T_{o}}[y\mid \mathbf{z}^{*}_u, \Theta, \mathbf{\pi}] = \sum_{p \in \mathcal{P}_o}\pi_{p_y} (\prod_{d \in \mathcal{D}} f_{d}(\mathbf{z}_u^{*};\Theta)^{\Gamma_{left}}\overline{f}_{d}(\mathbf{z}_u^{*};\Theta)^{\Gamma_{right}})
\end{equation}
where $\pi_{p_y}$ denotes the probability of a user with label $y$ reaching prediction node $p$, and $\overline{f}_{d}(\mathbf{z}^{*}_u;\Theta) = 1 -  f_{d}(\mathbf{z}^{*}_u;\Theta)$. To utilize the decision function of a decision node $d$ for explicitly routing the input to either the left or right, we introduce two binary indicators. Here, $\Gamma_{left}$ is $1$ if $p$ is the left subtree of node $d$, while $\Gamma_{right}$ is $1$ if $p$ is the right subtree of node $d$. Otherwise, they will be $0$.

Then the forest of decision trees is essentially an ensemble of decision trees $\mathcal{T} = \{T_{1}, ..., T_{O} \}$, which delivers a prediction with the given input $\mathbf{z}_u^*$ by averaging the output of each tree:
\begin{equation}\label{eq:PT}
    \mathbb{P}_{\mathcal{T}}[y\mid \mathbf{z}^{*}_u,\Theta,\mathbf{\pi}] = \frac{1}{O}\sum_{o=1}^{O}\mathbb{P}_{T_{o}}[y \mid \mathbf{z}^{*}_u,\Theta,\mathbf{\pi}]
\end{equation}
Hence, the predicted label $\hat{y}_u$ for user $u$ would be:
\begin{equation}
    \hat{y}_u =\underset{y}{argmax} \mathbb{P}_{\mathcal{T}}[y\mid \mathbf{z}^{*}_u,\Theta,\mathbf{\pi}], \,\,\,y\in \{0,1\}
\end{equation}
where $u$ is predicted as a genuine user if $\hat{y}=0$, and a fraudster if $\hat{y}=1$.

\subsection{Model Training}
In this section, we define the loss function of GraphRfi for model training. For rating prediction, as it is formulated as a regression task, we can quantify the prediction loss using squared error, i.e., $\sum_{\forall u,v\in \mathcal{E}}(r_{uv}-\hat{r}_{uv})^2$. However, in real-world scenarios, the observed ratings inevitably contain untruthful ones that are recognized as shilling attacks. Such untruthful ratings are extremely unreliable as they are published with malicious intent (e.g., hyping or defaming an item) to cheat the recommender systems. Meanwhile, directly optimizing GraphRfi through least squares which has an unbounded influence function is non-robust~\cite{mehta2007robust}. To enhance the robustness of our model, we additionally take $\mathbb{P}_{T}[y=0\mid \mathbf{z}^{*}_u, \Theta, \mathbf{\pi}]$, i.e., user $u$'s probability of being a genuine user (Eq.(\ref{eq:PT})) as bounded influence function. If the probability is low, GraphRfi can reduce the corresponding contribution of this suspicious user to the rating prediction task, as the rating given by this user is intrinsically unfair. To this end, the rating prediction loss is formulated as:
\begin{equation}
	\mathcal{L}_{rating} = \frac{1}{|\mathcal{E}|}\sum_{\forall u,v \in \mathcal{E}}\mathbb{P}_{T}[y=0\mid \mathbf{z}^{*}_u, \Theta, \mathbf{\pi}] \cdot (r_{uv} ^{'}-r_{uv})^{2}
\end{equation}
At the same time, for fraudster detection, we adopt cross-entropy, which is a widely adopted loss function for binary classification:
\begin{equation}
    \mathcal{L}_{fraudster} = \frac{1}{|\mathcal{U}|} \sum_{\forall u \in \mathcal{U}, y_u \in \mathcal{Y}} -\log \mathbb{P}_{T}[y= y_u\mid \mathbf{z}^{*}_u, \Theta, \mathbf{\pi}]
\end{equation}

Instead of training these two tasks separately, we combine their losses and jointly minimize the following loss function:               
\begin{equation}
    \mathcal{L} = \mathcal{L}_{rating} + \lambda \mathcal{L}_{fraudster}
\end{equation}
where $\lambda$ is a hyper-parameter to balance the effect of both parts. On one hand, the predicted probability in Eq.(\ref{eq:PT}) is utilized by the rating prediction component to guide the optimization of $L_{rating}$.
On the other hand, the predicted ratings will be further treated as an auxiliary feature in fraudster detection, of which the performance can be optimized via $L_{fraudster}$. Hence, both tasks are closely hinged with each other in our jointly training scheme.

\section{EXPERIMENTS}
In this section, we conduct experiments on two real-world datasets to evaluate the performance of GraphRfi on fraudster detection and robust rating prediction. Particularly, we aim to answer the following research questions (RQs):
\begin{itemize}
    \item \textbf{RQ1:} How does GraphRfi perform in the rating prediction task without/with the existence of fraudsters compared with baseline methods?
    \item \textbf{RQ2:} How does GraphRfi perform when detecting different types of fraudsters compared with baseline methods?
    \item \textbf{RQ3:} How is the recommendation robustness of GraphRfi w.r.t. each specific type of the shilling attack?
    \item \textbf{RQ4:} How do the hyper-parameters affect the performance of GraphRfi in different tasks?
\end{itemize}
\vspace{-0.5em}
\subsection{Experimental Datasets}
To simultaneously validate the performance of GraphRfi on both tasks, we consider two large-scale datasets, namely Yelp and Movies \& TV. Table~\ref{tab:statistic} shows the main statistics of these two datasets. We describe their properties below:
\begin{table}
\centering
  \scalebox{1.0}{%
  \begin{tabular}{l|l|l}
  \hline
    Network&Yelp&Movies\&Tv\\
    \toprule
    \#Users&32393&12630\\
    \hline
    (\%genuine users, \%fraudsters)&(70\%, 30\%)&(70\%,30\%)\\
    \hline
    \#Products&4670&4746\\
   \hline
   \#Ratings&293936&250423\\
   \hline
  \end{tabular}}
  \caption{Basic statistics of Yelp and Movies \& Tv.}
\vspace{-3em}
  \label{tab:statistic}
\end{table}

\textbf{Yelp}: Yelp has a filtering algorithm in place that identifies fake reviews and credible reviews. In our experiments, we use genuine reviews and fake reviews collected by~\cite{DBLP:conf/sigkdd/Akoglu15} to label normal users and fraudsters. Specifically, users who have posted fake reviews are labeled as fraudsters, while users with none fake reviews are marked as genuine users.

\textbf{Movies \& TV}: This dataset contains a series of users' ratings about movies and TV crawled from Amazon by~\cite{mcauley2013amateurs}. Following \cite{kumar2018rev2}, we use the helpfulness votes associated with each user's reviews for labelling normal users and fraudsters. Specifically, we pick users who received at least 20 votes in total. Then, a user is benign if the proportion of helpful votes is higher than $0.7$, and fraudulent if it is below $0.3$.  
\subsection{Evaluation Protocols}
To evaluate the rating prediction accuracy, we adopt two popular metrics, i.e., Mean Absolute Error (MAE) and Root Mean Square Error (RMSE). Smaller values of MAE and RMSE indicate better accuracy. For recommender systems, the existence of fraudsters is likely to affect the accuracy of predicted ratings. Thus, to validate the robustness of different recommenders, we start testing with models purely trained on genuine ratings, then infuse fraudsters into the training set and observe the performance fluctuations of all models. We randomly choose $20\%$ of the genuine ratings as the test set, and the remaining genuine ratings as the initial training set. Afterwards, we gradually enlarge the training set with $20\%$, $40\%$, $60\%$, $80\%$ and $100\%$ of the fraudsters and corresponding ratings.

For fraudster detection, we leverage widely-used classification metrics \cite{kumar2018rev2,dong2018opinion} Precision, Recall and F1 Score to evaluate the performance. We randomly choose $80\%$ of the labelled users as the training set and use the remaining labelled users as the test set.
\vspace{-0.5em}   
\subsection{Parameter Settings}
In GraphRfi, the ensemble size of decision trees is fixed to $5$, each with a depth of $3$. For the embedding size $e$, we tested the value of $\{50, 100, 150, 200, 250\}$, and the hyper-parameter of $\lambda$ was searched in $\{1, 3, 5, 7, 9\}$. Besides, we set the size of the hidden layer as $100$ and the activation function as $ReLU$ for the rating prediction component. Note that we adopt three-layer networks for all the neural components and model parameters were initialized with a Gaussian distribution, where the mean and standard deviation are $0$ and $0.1$. For all baseline methods, we use the optimal hyper-parameters provided in the original papers.                                            \vspace{-0.5em}                                          
\subsection{Baselines}
\begin{figure*}[t!]
\centering
\begin{tabular}{cccc}
	\multicolumn{4}{c}{\hspace{0.5cm}\includegraphics[scale=0.68]{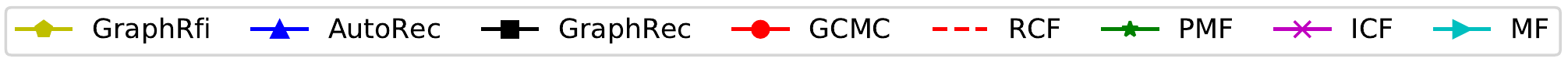}}\\
	\includegraphics[width=1.60in]{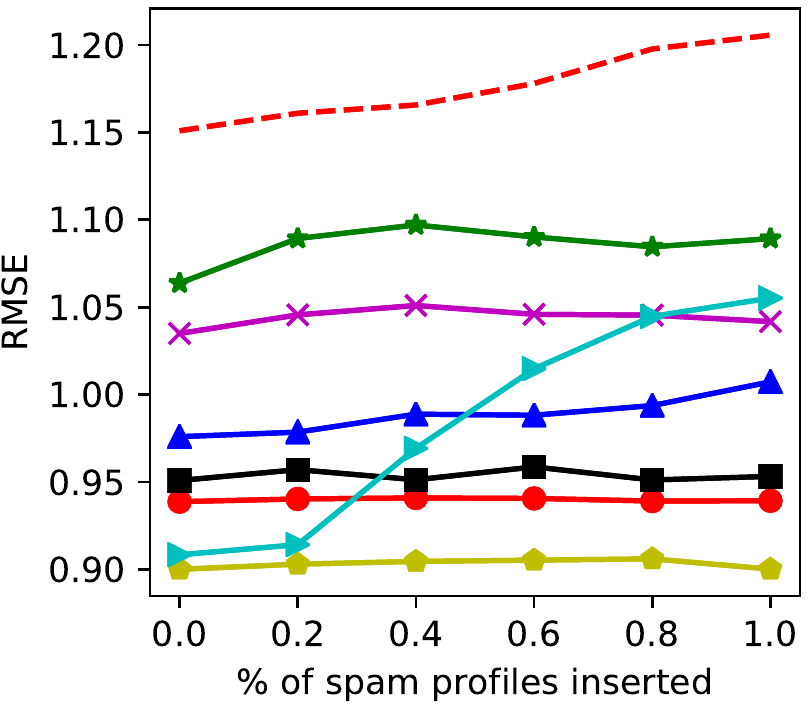}
	&\includegraphics[width=1.59in]{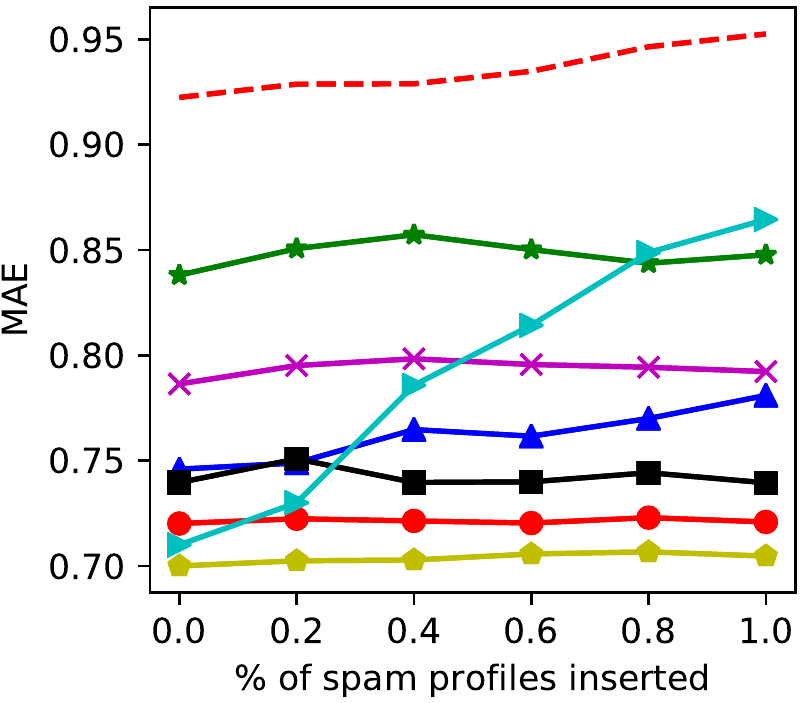}
	&\includegraphics[width=1.6in]{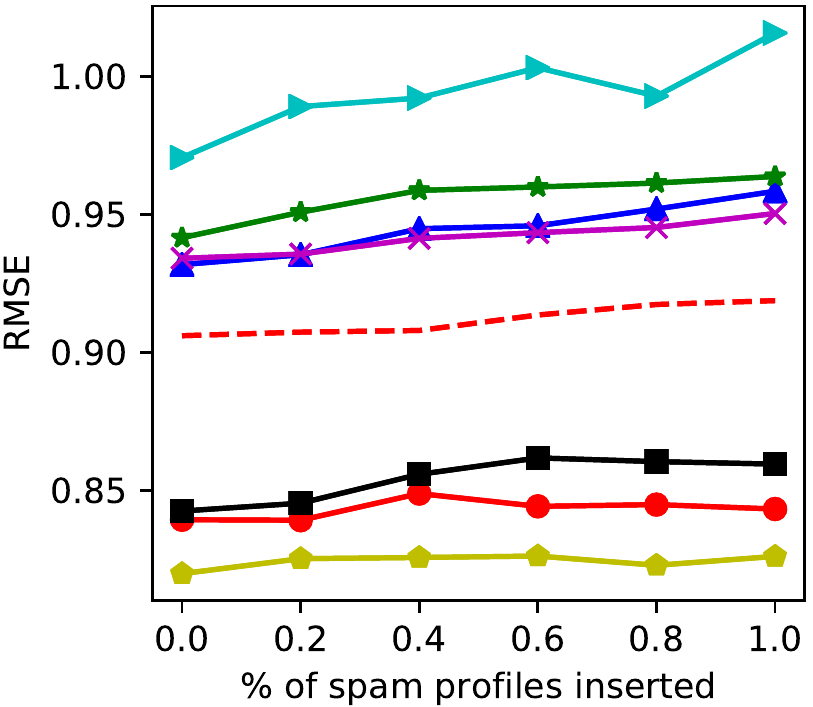}
	&\includegraphics[width=1.59in]{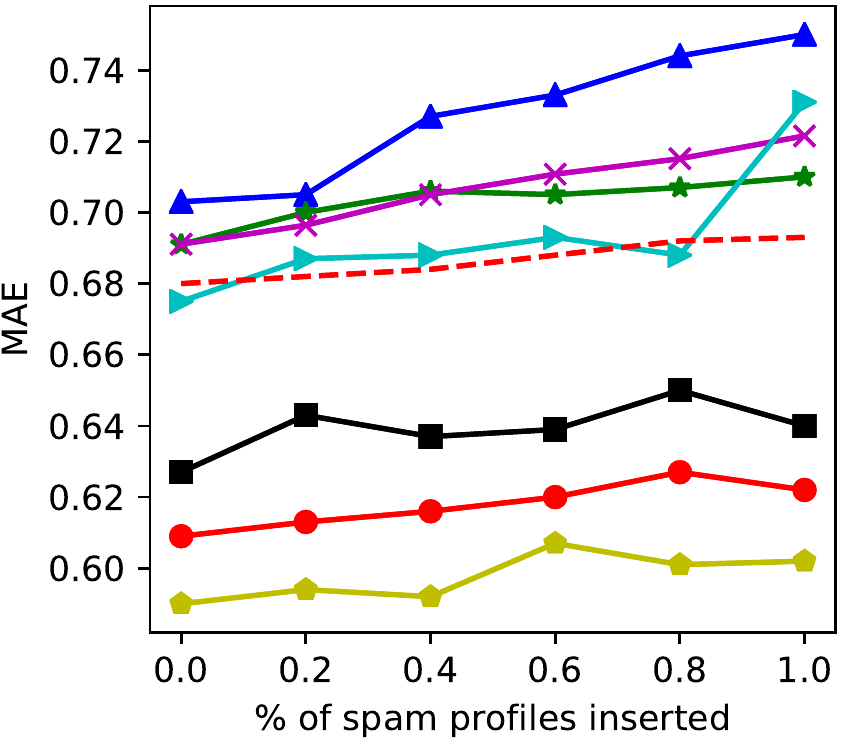}\\
	\hspace{0.6cm}\small{(a) RMSE on Yelp}
	&\hspace{0.6cm}\small{(b) MAE on Yelp}
	&\hspace{0.6cm}\small{(c) RMSE on Movies \& TV}
	&\hspace{0.6cm}\small{(d) MAE on Movies \& TV}
		\end{tabular}
\vspace{-0.4cm}
\caption{Rating prediction results on Yelp and Movies \& TV. Spam profiles refer to fraudsters in datasets.}
\vspace{-0.7em}   
\label{fig:rr}
\end{figure*}

\begin{table*}
\centering
  \scalebox{0.84}{%
  \begin{tabular}{|c|c|c|c|c||c|c|c||c|c|c||c|c|c|}
    \hline
     \multirow{2}{*}{Dataset}&\multirow{2}{*}{Method}&\multicolumn{3}{c||}{Mixed Attack}&\multicolumn{3}{c||}{Hate Attack}&\multicolumn{3}{c||}{Average Attack}&\multicolumn{3}{c|}{Random Attack}\\
    \cline{3-14}
   & &Precision&Recall&F1 Score&Precision&Recall&F1 Score&Precision&Recall&F1 Score&Precision&Recall&F1 Score\\
    \hline 
    \hline
   \multirow{5}{*}{Yelp}&Rev2&0.783&0.680&0.730&0.681&0.941&0.790&0.874&0.869&0.871&0.908& 0.935&0.921\\
   & DegreeSAD&0.923&0.918&0.921&0.926&0.919&0.922&0.922&0.917&0.919&0.922&0.920&0.921\\
   & FAP&0.688&0.013&0.026&0.796&0.015&0.030&0.742&0.014&0.028&0.711&0.014&0.027\\
   & OFD&0.941&0.991&0.966&0.935&\textbf{0.998}&0.966&0.983&0.982&0.982&0.979&\textbf{0.993}&0.986\\
   & GraphRFI&\textbf{0.989}&\textbf{0.992}&\textbf{0.990}&\textbf{0.979}&0.995&\textbf{0.987}&\textbf{0.993}&\textbf{0.991}&\textbf{0.992} &\textbf{0.987}&0.992&\textbf{0.990}\\
   \hline
   \hline
    \multirow{5}{*}{Movie \& TV} &Rev2&0.554&0.565&0.560&0.820&0.876&0.847&0.779&0.941&0.852&0.571&0.671&0.617\\
   & DegreeSAD&0.636&0.631&0.633&0.627&0.619&0.623&0.624&0.626&0.625&0.626&0.607&0.615\\
   & FAP&0.527&0.249&0.338&0.515&0.246&0.333&0.535&0.258&0.349&0.557&0.266&0.360\\
   & OFD&0.940&0.981&0.960&0.945&0.983&0.964&0.977&0.980&0.975&0.954&0.985&0.969\\
   & GraphRFI&\textbf{0.967}&\textbf{0.986}&\textbf{0.976}&\textbf{0.965}&\textbf{0.989}&\textbf{0.977}&\textbf{0.978}&\textbf{0.993}&\textbf{0.985}&\textbf{0.963}&\textbf{0.996}&\textbf{0.979}\\
   \hline   
  \end{tabular}}
  \caption{Fraudster Detection Performance on Yelp and Movies \& TV.}
\vspace{-2.5em}
  \label{tab:fd}
\end{table*}

We evaluate the rating prediction performance of GraphRfi by comparing with the following baselines:

\textbf{RCF}~\cite{mehta2007robust}: The Robust Collaborative Filtering model proposes a matrix factorization algorithm based on robust M-estimators against shilling attacks.

 \textbf{GCMC}~\cite{berg2017graph}: GCMC proposes a graph auto-encoder framework based on message passing on the bipartite interaction graph.
 
 \textbf{GraphRec}~\cite{wen2019www}: In this work, a novel graph neural network-based framework is proposed for social recommendation. As user-user interactions are unavailable in our case, we only use its user-item modeling component.
 
\textbf{AutoRec}~\cite{sedhain2015autorec}: This is a recent recommendation model that fuses collaborative filtering with autoencoders.

\textbf{PMF}~\cite{mnih2008probabilistic}: Probabilistic Matrix Factorization models the conditional probability of latent factors given the observed ratings and includes Gaussian priors to handle complexity regularization.

 \textbf{ICF}~\cite{sarwar2001item}: It stands for the Item-based Collaborative Filtering.
 
\textbf{MF}~\cite{koren2009matrix}: This is the Matrix Factorization algorithm that works by decomposing the rating matrix into the product of two lower dimensional user and item matrices.

Meanwhile, we compare GraphRfi with the corresponding fraudster detection models as follows:

\textbf{OFD}~\cite{dong2018opinion}: OFD is an end-to-end trainable model leveraging the properties of autoencoder and deep neural random forest. It is also the state-of-the-art method for fraudster detection.

\textbf{REV2}~\cite{kumar2018rev2}: REV2 proposes an iterative method to calculate three metrics (i.e., fairness, goodness and reliability) via the user-item rating matrix for distinguishing malicious users from genuine ones. 

\textbf{DegreeSAD}~\cite{li2016shilling}: It incorporates features relevant to item popularity to help detect fraudsters.

\textbf{FAP}~\cite{zhang2015catch}: It adopts a recursive bipartite propagation method to estimate each user's probability of being a fraudster.

\subsection{Prediction Performance (RQ1)}
\begin{figure*}[t!]
\centering
\begin{tabular}{cccc}
	\multicolumn{4}{c}{\hspace{0.5cm}\includegraphics[scale=0.68]{title.pdf}}\\
	\includegraphics[width=1.60in]{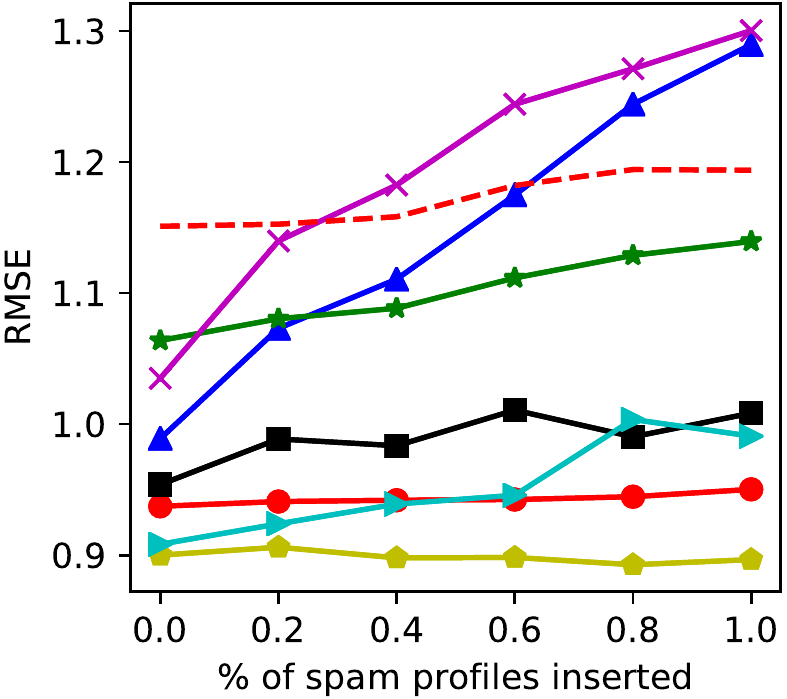}
	&\includegraphics[width=1.58in]{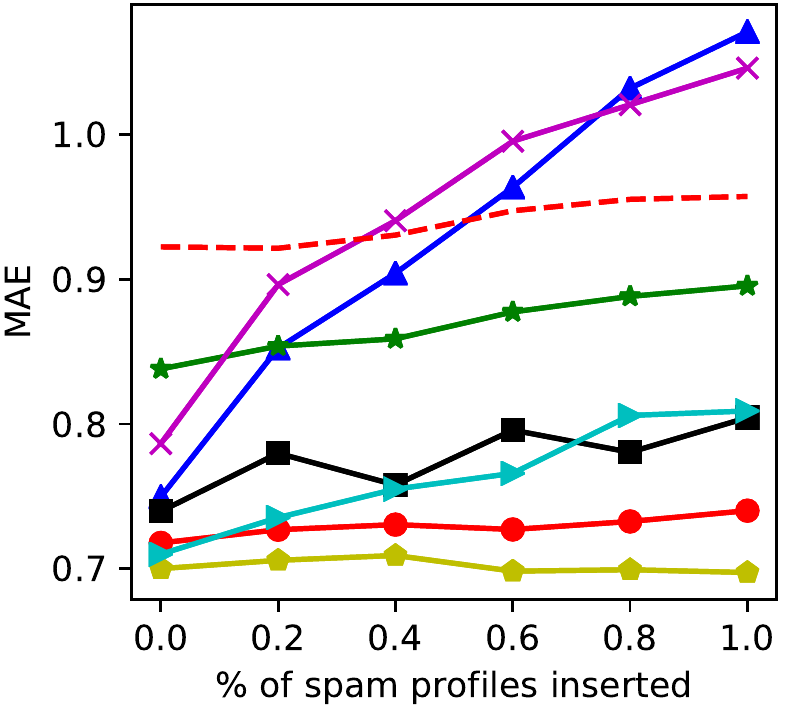}
	&\hspace{-0.2cm}\includegraphics[width=1.64in]{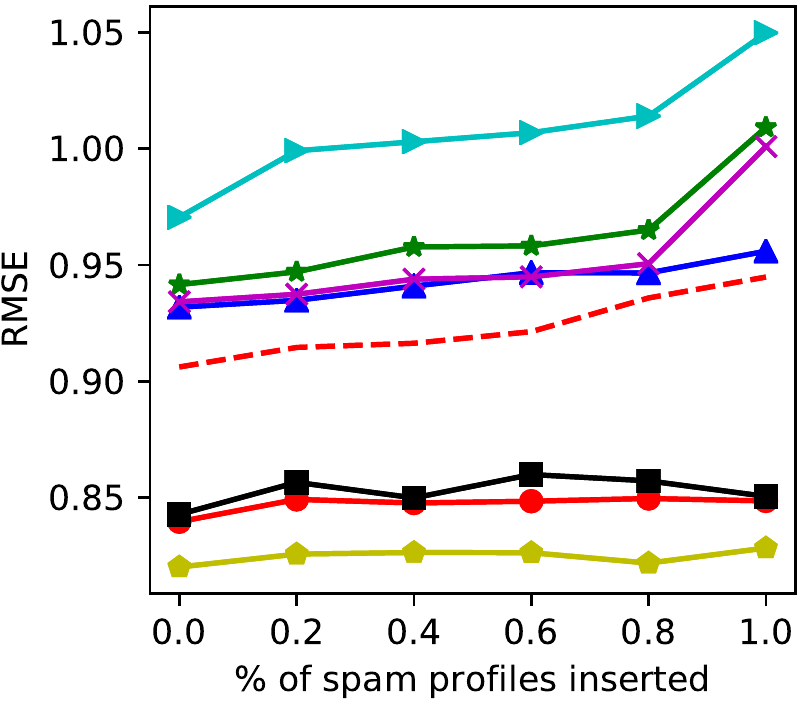}
	&\hspace{-0.4cm}\includegraphics[width=1.63in]{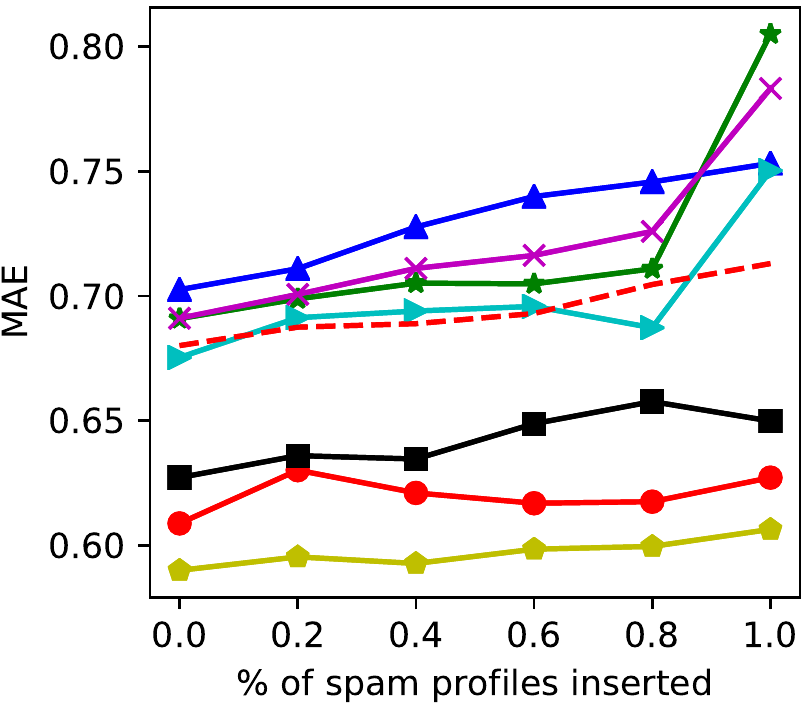}\\
	\vspace{-0.1em}
	\hspace{0.6cm}\small{(a) RMSE on Yelp - Random}
	&\hspace{0.6cm}\small{(b) MAE on Yelp - Random}
	&\hspace{0.1cm}\small{(c) RMSE on Movies \& TV - Random}
	&\hspace{0.1cm}\small{(d) MAE on Movies \& TV - Random}\\\vspace{-0.05em}
	\includegraphics[width=1.6in]{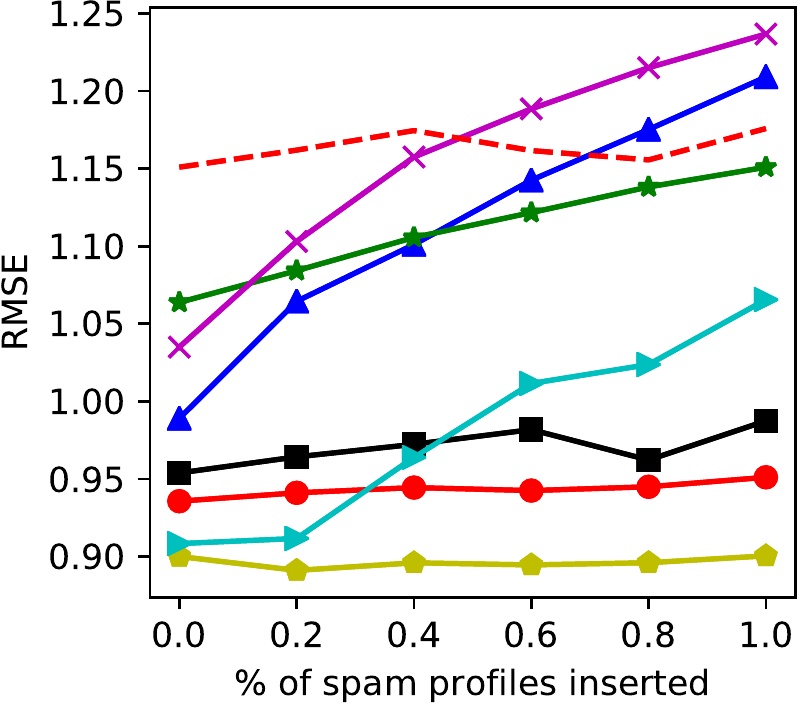}
	&\includegraphics[width=1.6in]{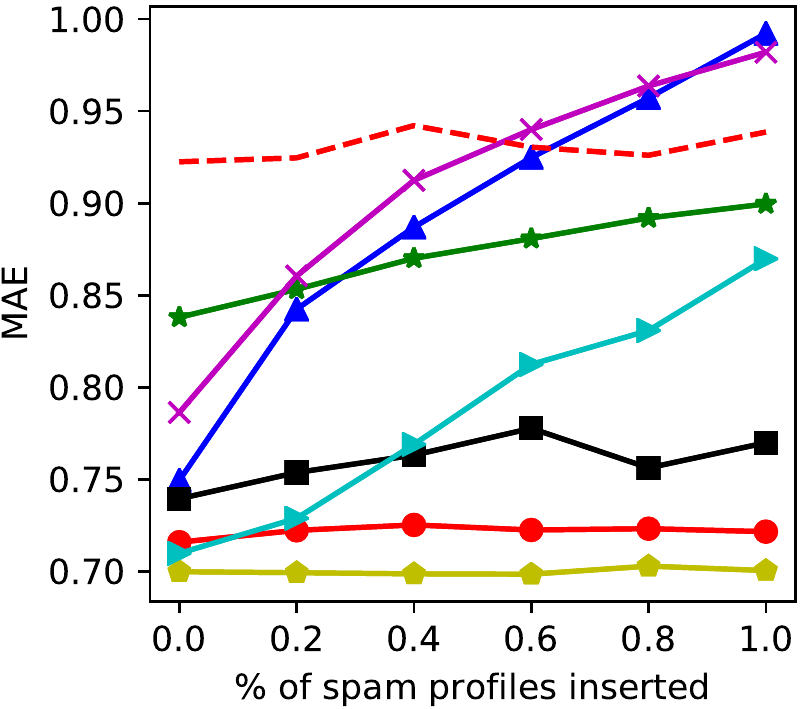}
	&\hspace{-0.2cm}\includegraphics[width=1.61in]{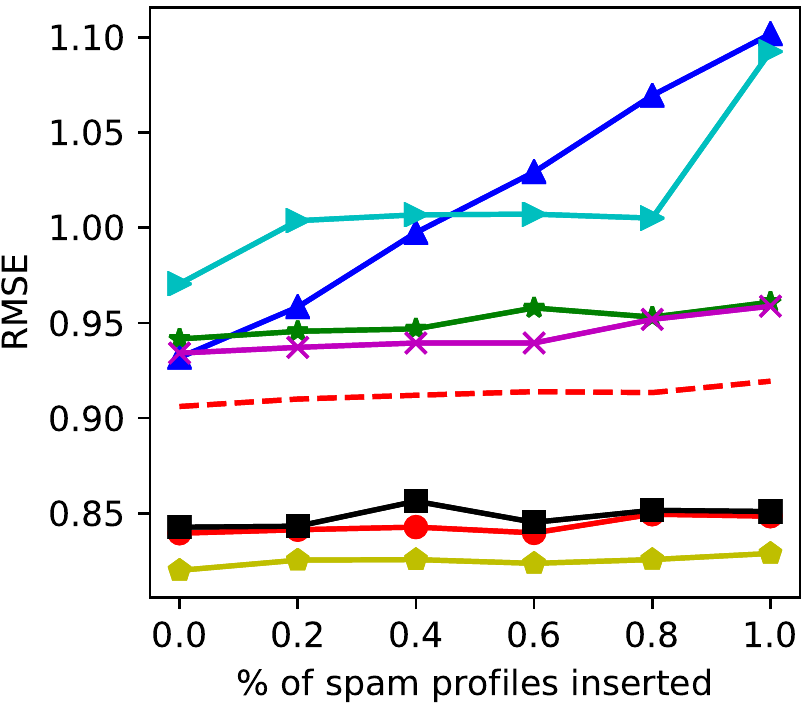}
	&\hspace{-0.4cm}\includegraphics[width=1.61in]{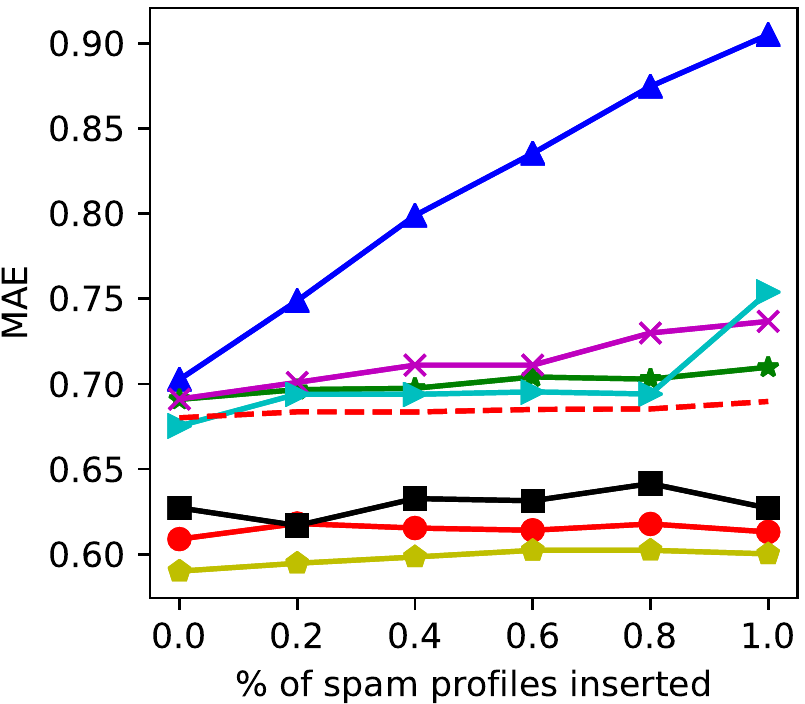}\\\vspace{-0.05em}
	\hspace{0.6cm}\small{(a) RMSE on Yelp - Average}
	&\hspace{0.6cm}\small{(b) MAE on Yelp - Average}
	&\hspace{0.1cm}\small{(c) RMSE on Movies \& TV - Average}
	&\hspace{0.1cm}\small{(d) MAE on Movies \& TV - Average}\\\vspace{-0.05em}
		\includegraphics[width=1.58in]{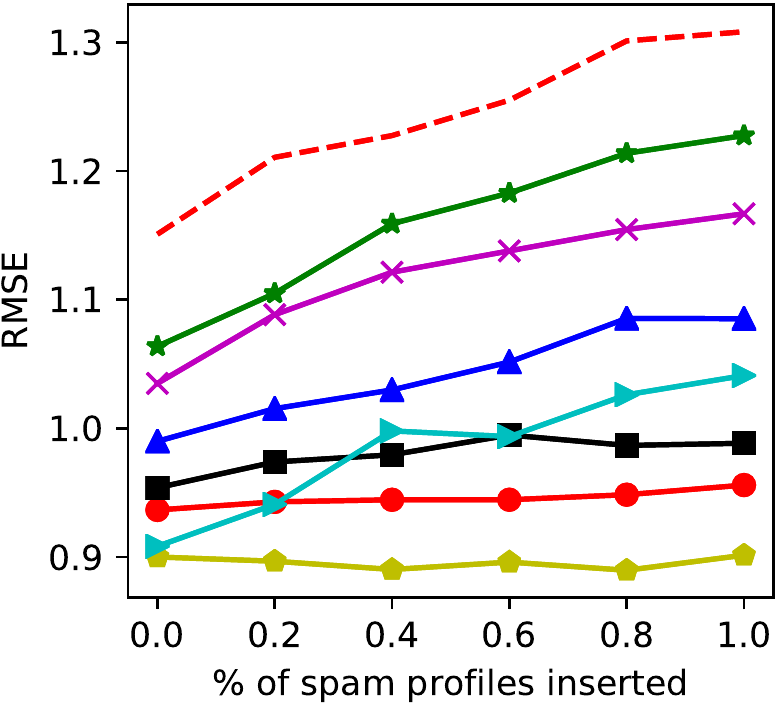}
	&\includegraphics[width=1.63in]{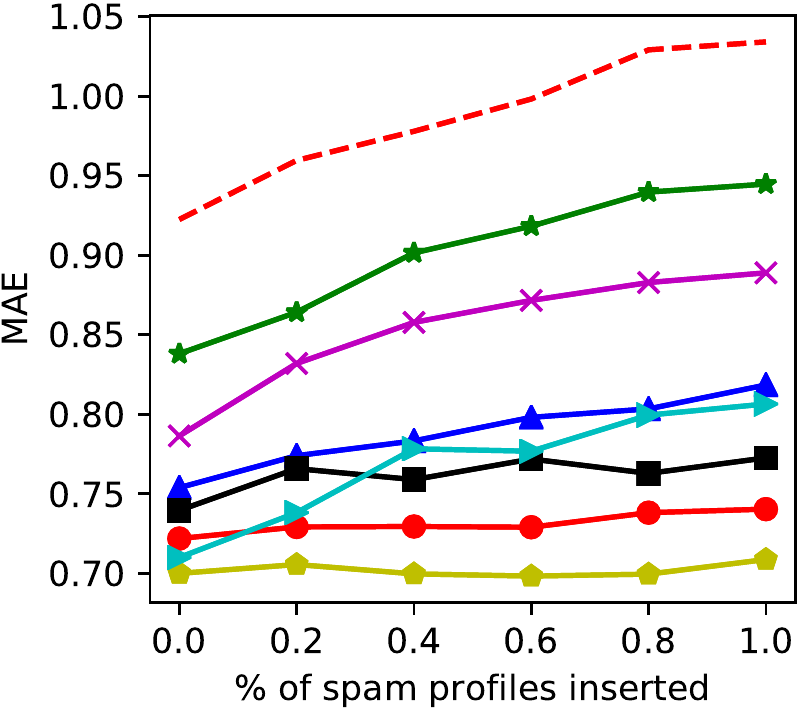}
	&\hspace{-0.2cm}\includegraphics[width=1.61in]{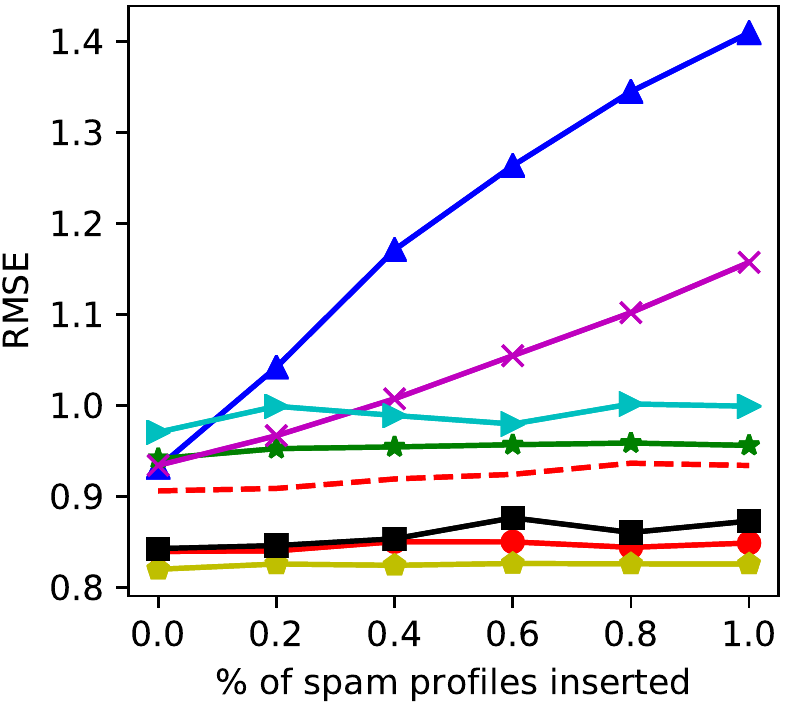}
	&\hspace{-0.4cm}\includegraphics[width=1.59in]{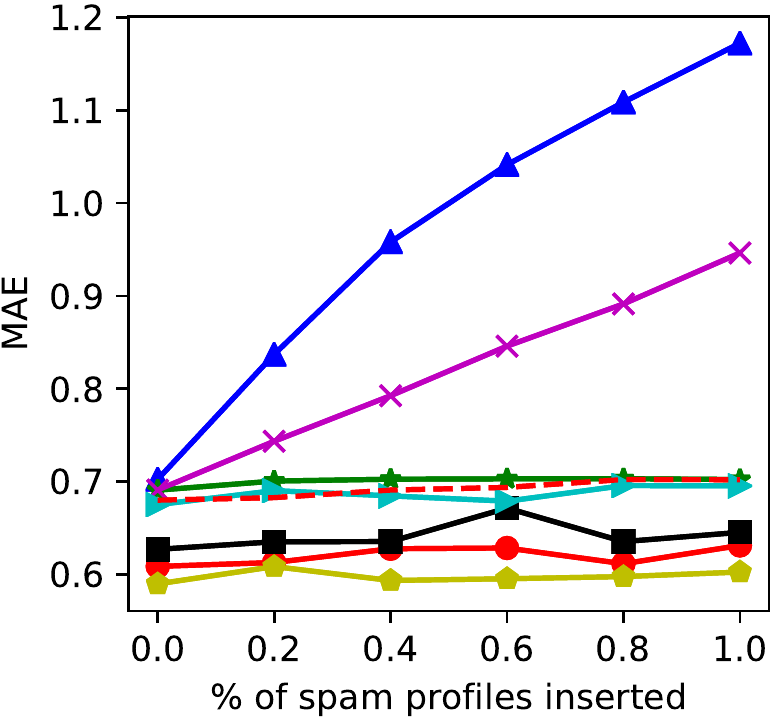}\\\vspace{-0.05em}
	\hspace{0.6cm}\small{(a) RMSE on Yelp - Hate}
	&\hspace{0.6cm}\small{(b) MAE on Yelp - Hate}
	&\hspace{0.1cm}\small{(c) RMSE on Movies \& TV - Hate}
	&\hspace{0.1cm}\small{(d) MAE on Movies \& TV - Hate}\\
		\end{tabular}
\vspace{-0.4cm}
\caption{Recommendation robustness on Yelp and Movies \& TV w.r.t. different types of shilling attacks. Spam profiles refer to fraudsters in datasets.}
\vspace{-1.8em}
\label{fig:rate}
\end{figure*}
Figure~\ref{fig:rr} shows the rating prediction error w.r.t. RMSE and MAE among all the recommendation algorithms. Clearly, our proposed GraphRfi significantly and consistently outperforms all baselines, even under the shilling attacks. For RMSE, we take Yelp as an example. Compared with MF which is the best baseline on Yelp when there are no fraudsters, GraphRfi has brought $0.8\%$, $1.1\%$, $6.5\%$, $10.9\%$, $13.9\%$ and $15.5\%$ improvements with the growing number of spam profiles inserted. Besides, GCN-based models (GraphRec and GCMC) generally perform better than other comparison methods on both datasets, indicating the superiority of GCN since it is able to preserve topology structure of a rating network as well as the node features. Furthermore, there is no constant winner among recommender models that are not based on GCNs. For example, the best performance among the five non-GCN models is achieved by MF with $0\%$, $20\%$ and $40\%$ fraudsters on the Yelp dataset, while AutoRec yields the best performance when the proportion of fraudsters are $60\%$, $80\%$ and $100\%$. RCF outperforms the other four non-GCN models on Movies \& and TV. Finally, although methods like ICF and PMF remain relatively robust in the presence of shilling attacks, they perform poorly in the rating prediction task on two datasets due to limited model expressiveness. 
\vspace{-1em}
\subsection{Fraudster Detection (RQ2)}
\label{sec:RQ2}
Driven by economic benefits, shilling attacks are becoming more common and diverse on e-commerce platforms. Generally, there are three widely agreed attack types in the relevant literature, namely random attack, average attack and hate attack~\cite{aggarwal2016recommender,DBLP:journals/air/GunesKBP14}. Usually, attacking only the target item cannot cause strong biases to the data and the recommender system, so fraudsters tend to post ratings on one target item and other filler items along with it, thus exerting more significant impact to the recommender systems. Target items are the set of items to be hyped or defamed, while filler items are the additional items to be injected, e.g., promoting a target item by giving low ratings to filler items and a high rating to it. As the original Yelp and Movies \& TV datasets are essentially a mixture of genuine ratings and different shilling attacks, we directly use them to test the overall fraudster detection performance of all models. To further test each model's classification accuracy w.r.t. each attack type, we follow \cite{aggarwal2016recommender} to generate simulated type-specific shilling attacks based on identified fraudulent users. Because each type of shilling attack reflects particular behavioral (i.e., rating) patterns, we mimic each type of shilling attack \cite{aggarwal2016recommender} by manipulating the ratings given by fraudsters. When detecting fraudsters with type-specific shilling attacks, for each fraudster, we choose the most popular item from her/his rated items as the target item, and all other rated items are treated as filler items. In the random attack, the filler items are randomly selected and assigned ratings based on a probability distribution w.r.t. the mean of all ratings, and the target item is assigned the maximum or minimum rating value, depending on whether it is a push or nuke attack. The average attack is similar to the random attack, and the only difference is that the filler items are assigned their average ratings in the dataset. The hate attack is specifically for the nuke attack, so the target item is assigned the minimum rating value, whereas the filler items are set to the maximum rating value.  

The results of the fraudster detection task are shown in Table~\ref{tab:fd}. We first conducted experiments on original datasets with the mixture of all attack types to test the overall performance in real-life scenarios, and then use the modified type-specific datasets for further evaluation. Obviously, on the original datasets, GraphRfi outperforms all the comparison methods by a large margin. Note that higher results imply better fraudster detection accuracy, which can be beneficial for blocking malicious reviews and helping online users make unbiased purchase decisions. Despite the slightly lower Recall of our model on Yelp-Random and Yelp-Hate compared with OFD, GraphRfi yields significantly higher performance in terms of Precision and F1 Scores. 
Correspondingly, it is evidenced that GraphRfi can preserve both users' behavioral features (defined in Table~\ref{tab:ff}) and graph structural patterns for fraudster identification. Besides, on both datasets, compared with the third best model DegreeSAD, the second best OFD brings $16\%$, $21\%$ and $18\%$ average improvement on Precision, Recall and F1 Score, respectively. This verifies the advantages of utilizing neural decision trees as the key building block for fraudster detection. In contrast to OFD, GraphRfi additionally utilizes each user's aggregated rating prediction error as an external feature source, thus leading to advantageous performance against OFD in most cases. In addition, FAP consistently underperforms, and a possible reason is 
that FAP does not make full use of users' behavioral features, which can cause immense information loss. In summary, the underlined superiority of GraphRfi proves that it is able to accurately detect fraudsters either with type-specific shilling attacks or in the real-life mixture scenarios.
\subsection{Robustness Performance (RQ3)}
\begin{figure*}[t!]
\centering
\begin{tabular}{cccc}
	\includegraphics[width=1.63in]{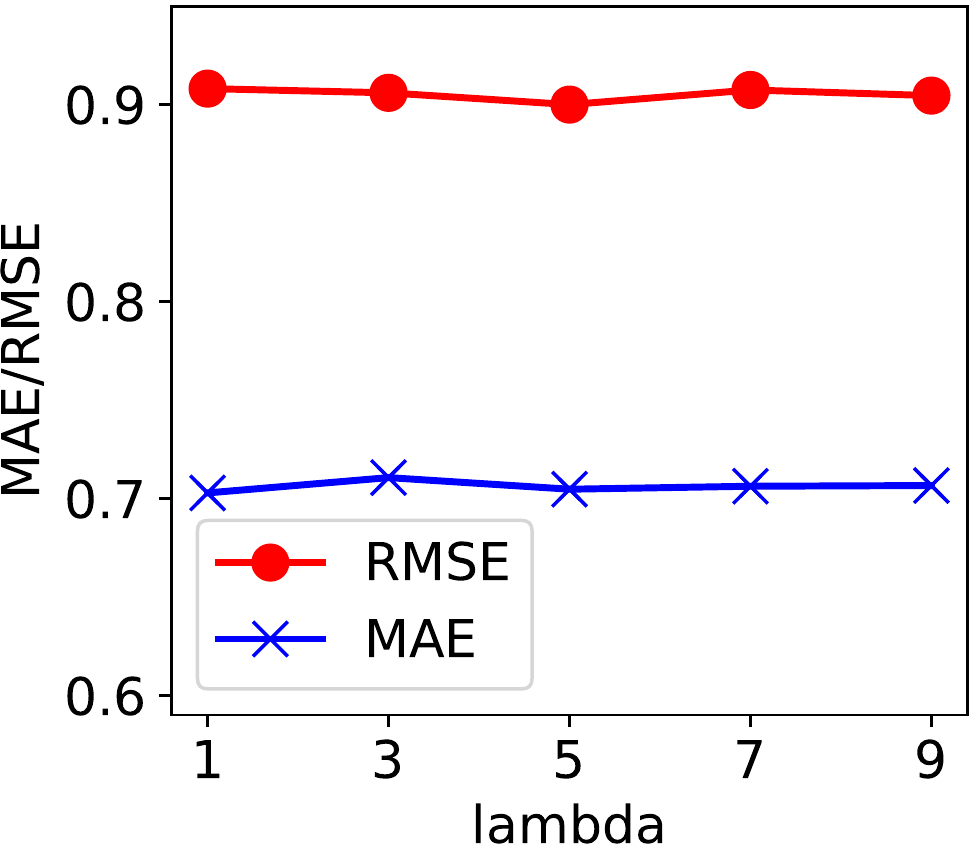}
	&\includegraphics[width=1.48in]{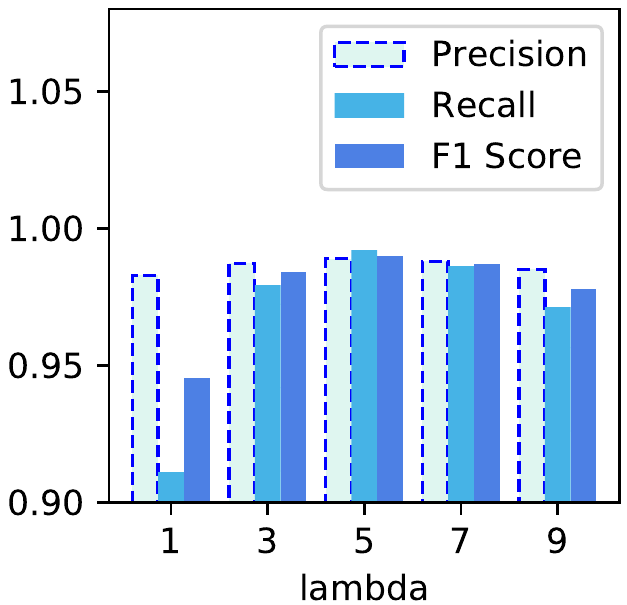}
	&\includegraphics[width=1.61in]{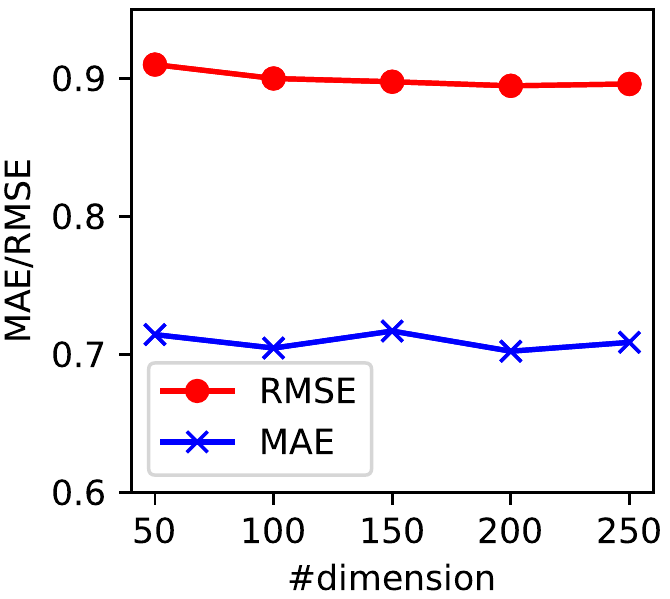}
	&\includegraphics[width=1.46in]{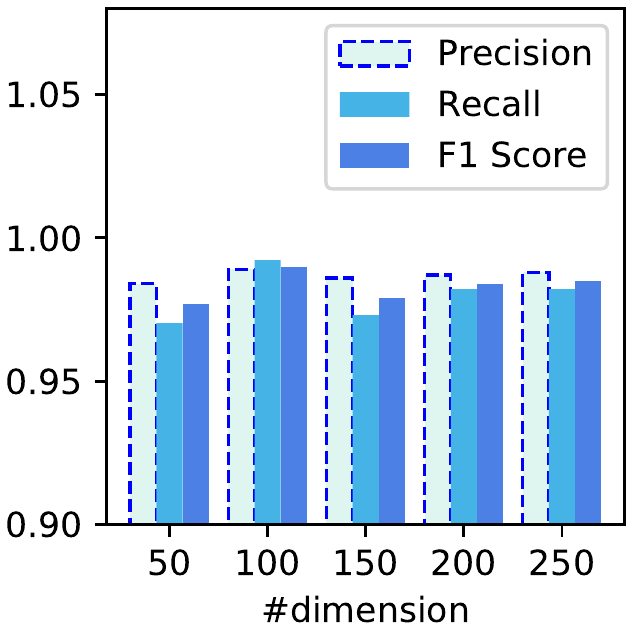}\\
    \hspace{0.6cm}\small{(a) Rating Prediction}
	&\hspace{0.6cm}\small{(b) Fraudster Detection}
	&\hspace{0.6cm}\small{(c) Rating Prediction}
	&\hspace{0.6cm}\small{(d) Fraudster Detection}
		\end{tabular}
\vspace{-0.4cm}
\caption{Parameter sensitivity w.r.t. $\lambda$ and dimension $e$ on Yelp.}
\label{fig:para}
\vspace{-1.2em}
\end{figure*}
Recommendation methods that show minor vulnerability to shilling attacks are referred to as robust recommenders. To answer RQ3, we investigate how our model's rating prediction component performs in the presence of aforementioned shilling attack types, namely random, average and hate attack. In this study, we train all recommenders on one type-specific dataset constructed in Section \ref{sec:RQ2} each time, and evaluate their rating prediction accuracy on the same test set. Figure~\ref{fig:rate} shows the influences of shilling attacks on the performance of all recommenders. We also vary the percentage of inserted fraudsters by $0\%$, $20\%$, $40\%$, $60\%$, $80\%$ and $100\%$.

From Figure~\ref{fig:rate}, we can see that our GraphRfi shows minimal impact from all types of attacks from fraudsters, and is the most robust model that generates consistently accurate predicted ratings. It is within our expectation that the high-performance NRF classifier in GraphRfi can effectively bond the impact of fraudsters, thus preventing the recommender from being corrupted by shilling attacks. Besides, the results w.r.t. different attack types show that hybrid models are generally more robust against attacks, while traditional collaborative filtering-based approaches are highly sensitive to fraudulent ratings. Noticeably, as two GCN-based methods, the performance fluctuations of GraphRec and GCMC are less severe than other methods as the number of fraudsters increases. It is because their predicted ratings are determined by accounting for external domain knowledge (e.g., side information) on top of the collaborative filtering effect. 
 \vspace{-0.5em}
\subsection{Impact of Hyper-parameters (RQ4)}
To verify the impact of different hyper-parameters to GraphRfi, we conduct a set of experiments with varied hyper-parameters on Yelp dataset. Specifically, we study our model's sensitivity to the embedding dimension $e$ in $\{50, 100, 150, 200, 250\}$ and the value of $\lambda$ in $\{1, 3, 5, 7, 9\}$. We show the performance differences by plotting RMSE and MAE for rating prediction, while demonstrating Precision, Recall and F1 Score for fraudster detection. Figure~\ref{fig:para} presents the results with different parameter settings. As can be inferred from the figure, all the best rating prediction and fraudster detection results are achieved with $\lambda=5$. Thus, we set $\lambda = 5$ to achieve a balance between the accuracy of both rating prediction and fraudster detection. Meanwhile, since the value of the latent dimension $e$ is directly associated with our model's expressiveness, GraphRfi generally benefits from a relatively larger $e$ in rating prediction. However, the performance on fraudster detection tends to become worse when $e$ increases, as the excessive network parameters introduced by a large $e$ may lead to the problem of overfitting. 
 \vspace{-0.5em}
\section{RELATED WORK}
\textbf{Recommender System.} Recommender systems play a pivotal role in enhancing user experience and promote sales for e-commerce. The collaborative filtering algorithm is a widely adopted recommendation technique, which utilizes the activity history of similar users (i.e., user-based recommender system~\cite{zhao2010user}) or similar items (i.e., item-based recommender system~\cite{sarwar2001item}). Recent years have witnessed the tremendous success of the deep learning techniques (DL) in recommender system~\cite{zhang2019deep,chen2019exploiting,he2020lightgcn,wang2018streaming}. The DL-based recommender systems can be classified into different categories based on the types of employed deep learning techniques. Many existing recommendation models extends traditional recommendation methods by utilizing multi-layer perceptron such as NCF~\cite{he2017neural}, DeepFM~\cite{guo2017deepfm} and CML~\cite{hsieh2017collaborative}.~\cite{wu2016collaborative,sedhain2015autorec} consider the collaborative filtering from Autoencoder perspective, while ~\cite{berg2017graph,wen2019www, ying2018graph} utilize graph convolutional networks for feature extraction. These methods have shown satisfactory performance in item recommendation. However, they trust all the input data and treat them equally, which may be violated in the presence of shilling attacks. Therefore, robust recommender algorithms have been specifically proposed with attack resistance.  RCF~\cite{mehta2007robust} propose to modify the optimization function by combining M-estimators to improve the robustness of traditional matrix factorization. Noticeably, RCF cannot integrate the rich side information of the users and items. Another line of robust recommendation~\cite{mehta2007robust, o2004evaluation,mehta2009unsupervised} thwarts attacks by detecting fraudsters first, and then the recommendations can be generated after the removal of fraudsters from the dataset. But it is risky to remove genuine users, resulting in the negative effect on rating prediction~\cite{aggarwal2016recommender}. Motivated by this, we proposed GraphRfi that can generate stable recommendations in the presence of shilling attacks. 

\textbf{Fraudster Detection.} Online fraud such as creating fake feedbacks has become a common phenomenon in recent years~\cite{wu2015spammers}. In that case, understanding the reliability of online reviews is of great importance to help online users make wise purchase decisions. Two main branches of rating fraud detection approaches are often distinguished: behaviour-based fraudster detection and network-based fraudster detection. Behaviour-based models achieve these purpose by extracting features from the review text~\cite{sandulescu2015detecting,fayazi2015uncovering} or time entropy~\cite{xie2012review,minnich2015trueview}. Some researchers extend work by incorporating behavior patterns~\cite{rayana2015collective} or the probabilistic distribution of rating behavior~\cite{hooi2016birdnest}. A survey on behavior-based algorithms can be found in~\cite{jiang2016suspicious}. Network-based models solve the fraudster detection by iterative learning, belief propagation and node ranking techniques.~\cite{li2012robust,wang2012identify,wang2011review,mishra2011finding} propose iterative approaches to jointly calculate scores in the rating networks. The recent work is~\cite{kumar2018rev2}, which creates three metrics, namely, fairness, goodness and reliability to qualify users, items and ratings. A comprehensive survey of network-based fraudster detection approaches can be found in~\cite{akoglu2015graph}. 
\vspace{-1em}
\section{CONCLUSION}
In this paper, we propose an end-to-end model named GraphRfi to perform user representation learning, which is able to jointly optimize the two-level tasks, namely robust rating prediction and fraudster detection. In GraphRfi, the graph convolutional network precisely captures user preference and node side information, while the neural random forests achieves satisfying accuracy in fraudster detection. Additionally, the strict classifier of fraudsters is expected to discourage large shifts in the rating prediction process, while an accurate rating prediction system tends to be beneficial to the fraudster detection. The experimental results based on two real-world datasets demonstrate the effectiveness and practicality of GraphRfi over the state-of-the-art baselines on both tasks. 
\section{Acknowledgments}
The work has been supported by Australian Research Council (Grant
No.DP190101985, DP170103954 and FT200100825).
\bibliographystyle{ACM-Reference-Format}
\balance
\bibliography{sample-base}


\end{document}